# Green Information Technology as Administrative innovation - Organizational factors for successful implementation: Literature Review

*Research in progress*

**Badrunnesa Zaman**
Information Systems School, Queensland University of Technology
2 George Street, Brisbane, Australia, QLD-4000
Email: mayabi79@gmail.com

**Darshana Sedera**
Information Systems School, Queensland University of Technology
2 George Street, Brisbane, Australia, QLD-4000
Email: d.sedera@qut.edu.au

## Abstract

There is a considerable amount of awareness of environmental issues and corporate responsibility for sustainability. As such, from a technological viewpoint, Green IT has become an important topic in contemporary organizations. Consequently, organisations are expected to be innovative in their business practices to become more sustainable. Yet, the popularity and adoption of such initiatives amongst employees remain low. Furthermore, the management practices for adhering to Green IT are largely dormant, lacking active incentives for employees to engage in Green IT initiatives. This study observes the phenomenon of Green IT through administrative innovation. In doing so this paper performs a comprehensive analysis of 137 papers published between 2007 and 2015. The paper reveals organizational factors for successful implementation of Green IT as administrative innovation that can be useful to both academia and practice.

## 1  INTRODUCTION

Organisations fail to enact their Green IT (Information Technology) initiatives due to a lack of innovativeness in their approaches and for failing to gather momentum from the respected stakeholders (Baregheh, Rowley et al. 2009). Such initiatives have been largely ineffective due to organisations lacking understanding of their readiness to adopt the Green IT initiatives, the lack of organizational structures, and the unavailability of continuous sponsorship (Deng et al. 2009; O'Neill 2010). Additionally, the popularity and adoption of such initiatives amongst employees remain low. Furthermore, the management practices promoting the adherence to Green IT are largely dormant, lacking active incentives for employees to engage in Green IT initiatives. Considering the tough economic challenges the majority of organisations are currently facing due to the aftermath of the global financial crisis, Green IT may not be perceived to be a critical organizational priority (O'Neill 2010). Companies who are implementing Green IT can face barriers such as inadequate funding; inadequate skills to implement the initiatives; unclear objectives or goals; and inadequate system to support the technical necessities (Bose and Luo 2011).

It is widely recognised throughout literature and practice that hardware has an enormous detrimental impact on the environment. The illegal dumping of hardware may lead to serious health issues (Loeser 2013). Also, energy consumption by IT services (e.g. data centres) indirectly contribute to $CO_2$ emissions. Moreover the excessive use of IT in our everyday life has become the cause of several environmental issues including carbon dioxide ($CO_2$) emissions (Herzog et al. 2012) and potentially global warming (Hasan and Meloche 2013). Therefore, both IT hardware manufacturers and the organizations using IT need to apply environmental sustainability policies containing product recycling, pollution prevention, and sustainable development in managing IT (Molla and Abareshi 2011). According to the world commission on Environment and Development, sustainability is a main business competency that safeguards without harming the future generations (Angeles and Fredericton 2015).

To overcome the aforementioned problems, the IT industry can play a pivotal role by innovating to reduce the impact of IT on the environment, which is commonly known as Green IT. Green IT refers to the introduction of innovative IT products, application, services and practices that reduce the impact of IT on the environment (Murugesan 2008a). Adopting the definition of innovation by Thompson Green IT innovation can be defined as the generation, acceptance and implementation of new ideas, processes, products or services of IT that reduces environmental impact of IT (Thompson 1965) designed to benefits organisation's stakeholders (West and Anderson 1996). Organizations need to innovate for ongoing customer demands and in order to make the most from the technology (Baregheh et al. 2009). Consequently Green initiatives are gaining in prominence, and in some cases mandatory for organizations. We observe this phenomenon of Green IT through administrative innovation, where administrative innovation refers to introducing new procedures, policies and organizational forms (Lin and Zhang Zhao 2010).

In practice, organizations are continually devoting resources in an attempt to be innovative in their business practices to become more sustainable. Yet, green IT initiatives once implemented are usually unsuccessful and demand a deeper understanding. Furthermore there is a dearth of empirical studies devoted to the adoption of Green IT policies, practices and technologies that both improve the energy consumption of IT infrastructure and considers the environmental criteria in the procurement and disposal of IT (Kuo and Dick 2010a; Standing et al. 2008). Furthermore, according to Tan et al. (2011) there is a gap in understanding "*how green IT initiatives can be effectively implemented*". To overcome these gaps, this research-in-progress paper conducts a comprehensive archival analysis of 137 papers to uncover the organizational factors that are imperative for the successful implementation of Green IT. These factors will then be linked to the stages of the innovation process to enhance the understanding of Green IT as an administrative innovation. The findings from this literature review will help future researchers identify the management issues as well as the implementation issues related to Green IT. Additionally this review will provide guidance for the decision makers developing and implementing Green IT policies. Consequently, the insightful findings from this paper will be useful for both academics and practitioners.

This paper is organised as follows: next a brief review of the literature is presented to articulate the concept of Green IT as an administrative innovation. Subsequently, the research method is depicted, followed by the results of a comprehensive archival analysis. The paper concludes with areas for future research.

## 2   LITERATURE REVIEW

In order to perform the archival analysis, a comprehensive understanding of Green IT and Green IT as an administrative innovation was necessary. Consequently, this section seeks to provide a brief overview of the conceptual domain of Green IT as an Administrative Innovation.

### 2.1  What is Green IT

Recently, organizations have realized the long term effects of IT and therefore they have commenced taking initiatives to improve their environmental footprint (Molla 2009a). There are many view and definitions of Green IT. A plethora of Green IT definitions exist throughout the literature. However, the underlying themes of these definitions pertain to the energy intakes and the waste associated with the use of computer hardware and software (Angeles and Fredericton 2015).

Green IT is the systematic application of practices that enable the minimization of the environmental impact of IT, maximise efficiency and allow for company-wide emission reductions based on technology innovations (Elliot 2011; Erek et al. 2011). Furthermore, Murugesan (2008b defined Green IT as the study and practice of designing, manufacturing, using, and disposing of IT products such as computers, servers, monitors, printers, storage devices, and networking and communications systems - efficiently and effectively with minimal or no impact on the environment. Green IT is not only about the technology and environment; it also endeavours to achieve economic feasibility and improved system performance and use, while abiding by social, legal and ethical responsibilities within organisations (Elliot 2007; Murugesan 2008b).

For any organisations intending to improve its Green IT accreditation, they must first establish a Green IT policy including the design, manufacture, use and disposal of IT products (Murugesan 2008b) depending on the area they operate their business. Then all the stakeholders which may include its staff, clients, suppliers and users have to be made aware of where the organisation stands on Green and environmental issues related to IT. Additionally, the policy must be adopted, authorised and committed to the employees by the organisation's senior management (O'Flynn 2009). The Green IT must be aligned with their business strategies.

So, what are the fundamental requirements for any organisation considering going green? Setting up Green IT policies, and explaining these policies to the employees, dealers, clients and consumers (O'Neill 2010) are considered to be the starting point for any Green IT initiatives. There are many key factors for any Green IT initiatives to succeed. Once developed, the organisation must ensure that they are communicated and consulted with the stakeholders and have their 'buy in'. Further, Green IT should be aligned with the current organisational strategies and have administrators commitment to this initiative because organizations that, regardless of their size, considered environmental issues as a business and opportunity challenge have succeeded (Subburaj and Kulkarni 2014).

## 2.2 Green IT as an administrative innovation

We observe this phenomenon of Green IT through administrative innovation. Organizational modification is a crucial feature in the life cycle of a business. Introduction of new products, services, organizational forms are the preferred substitutes that organizations use to compete to satisfy their customers (Ettlie and Reza 1992). Hence, it is reasonable and logical to assume that a 'new' type of innovation would bring forward administrative innovation to the organization (Damanpour 1991, Green, Gavin et al. 1995). We consider Green IT as an administrative innovation because organisation may require "*to improve organizational structures and systems, administrative processes, management systems, and departmental coordination as well as recruitment and personnel policies, control and motivation systems and structuring of organizational activities*" (Damanpour 1987p. 677) to successfully adopt Green IT.

Innovation is an extensively used idea that is variously defined to reflect the specific necessity and features of a specific area of study (Damanpour and Evan 1984). The different contexts of innovation are very diverse. According to Damanpour "*Innovation is a means of changing an organization, whether as a response to changes in its internal or external environment or as a pre-emptive action taken to influence an environment*" (Damanpour 1991p. 556).

Moreover IT innovation is defined as the innovation in the firm's application of digital computer and communication technologies (Swanson 1994). Green IT includes a firm's application of new types of communication technologies which are aimed to reduce the environmental impact of a firm's operation. Therefore Green IT is considered to fall under the umbrella of IT innovation (Nishant et al. 2011). Green IT initiatives are new and needs more exploration (Nishant et al. 2011).

Many scholars have come up with the idea and definition of different kinds of innovation in different fields of study (Baregheh et al. 2009). As our main focus of the study is the organizational side of innovation we will look deeper into the organizational innovation. Organizational innovation is applying new and more improvised organizational methods for increasing organizational performance (OECD 2005). According to Daft's dual core model of innovation, organizational innovation can be of two types: i) Technological innovation and ii) Administrative innovation (Daft 1978).

According to Damanpour and Evan (1984, p. 560) "*Administrative innovations involve organizational structure and administrative processes, they are directly related to the basic work activities of an organization and are more directly related to its management*". And technological innovations "*are defined as innovations that occur in the technical system of an organization and are directly related to the primary work activity of the organization*"(Damanpour and Evan 1984p.394)

By reviewing the above discussion and considering different types of innovation definitions, Green IT innovation can be related with all these innovation types as organisation may require innovating new products, improving their processes, establishing new marketing strategies and will require being innovative in the way they operate their organisation (i.e. administration).

According to Damanpour (1996 innovation is "*A process that includes the generation, development, and implementation of new ideas or behaviours. Further, innovation is conceived as a means of changing an organization, either as a response to changes in the external environment or as a pre-emptive action to influence the environment. Hence innovation is here broadly defined to encompass a range of types, including new products or services, new process technologies, new organizational structures or administrative systems, or new plans or programs pertaining to organizational members*". Damanpour and Evan described administrative innovation "as those that occur in social system of an organization ...an administrative innovation can be implemented of a new way to recruit personnel, allocate resources and structure tasks, authority and rewards. It comprises innovation in organizational structure and in the management of people" (Damanpour and Evan 1984p. 394).

Therefore, by analysing the literature and innovation classifications as defined earlier we believe that Green IT is a new process which if introduced to the organization can lead to administrative innovation. This may lead to structural alignment and personnel changes and affect a wide dimension across all organizational levels and tasks (Sisaye 2015). Accordingly, requirement of new reporting systems, human resource policies, internal control mechanisms, resource allocation decisions, and cross-functional collaboration and coordination systems may arise (Sisaye 2015).

# 3 RESEARCH METHOD

We conducted an archival analysis on Green IT publications in the top Information Systems (IS) journals and conferences from the years 2007 to 2015.

Our research started from the senior scholar's basket of eight journals which are:

i. European Journal of Information Systems;
ii. Information Systems Journal;
iii. Information System Research;
iv. Journal for the Association of Information Systems;
v. Journal of Information Technology;
vi. Journal of Management Information Systems;
vii. Journal of Strategic Information Systems; and
viii. Management Information Systems Quarterly.

Due to the limited number of publications in the top tier outlets the search was extended other well-known IS and management journals such as Organizational Science, Information and Management, Academy of Management, Information Technology and People, Australasian Journal of Information Systems, Communications of the Association for Information Systems and the International Journal of Business management. We also searched the top conference proceedings such as ICIS, ACIS, ECIS, PACIS, and AMCIS[1]. The search included the keywords Green IT, Green IS, Green leadership, IT sustainability, Green Computer, Environmental sustainability and Green ICT. We started to search papers from the period of 2007 because from this year researchers started to look deeper into Green IT. According to Brooks et al. (2010 "Green IT" appeared for the first time in 2007 in a CIO magazine. We did not find any papers in the eight basket journals and also in major conferences regarding Green IT. Therefore we contemplate 2007 as the start our review.

In conducting our literature search, several academic databases and search engines were used such as AIS Electronic Library (Aisel), ACM Digital Library, Science Direct, Emerald, Google Scholar, Emerald Engineering Database, InfoSci collection and IEEE Xplore Digital Library. By searching these libraries, a total of 137 relevant papers were identified and used.

# 4 PUBLICATIONS PER YEAR

Over the past few decades the use of IT has increased in several areas such as organisations, governments, and civilizations and has ultimately improved the convenience of our lives across several dimensions (Murugesan 2008a). Table 1 depicts the total number of relevant articles reviewed from the journals and conferences on a year-by-year basis from 2007 to 2015. According to this table we can see that the number of publications have increased during the time period of 2007 to 2015.

| Year | Journal Article | Conference Article | Total |
|---|---|---|---|
| 2007 | 1 | - | 1 |
| 2008 | - | 5 | 5 |
| 2009 | 1 | 12 | 13 |
| 2010 | 5 | 18 | 23 |
| 2011 | 10 | 15 | 25 |
| 2012 | 3 | 15 | 18 |

---

[1] ICIS: International Conference of Information Systems, ACIS: Australasian Conference of Information Systems, ECIS: European Conference of Information Systems, PACIS: Pacific Asian Conference of Information Systems, AMCIS: Americas Conference of Information Systems

| | | | |
|---|---|---|---|
| 2013 | 9 | 9 | 18 |
| 2014 | 3 | 19 | 22 |
| 2015 | 2 | 10 | 12 |
| **Total** | **34** | **103** | **137** |

*Table 1: Distribution of Review Sample by publication year*

# 5   ANALYSIS, FINDINGS AND DISCUSSION

The topics related to environmental sustainability has become vital in Information System's research (Watson et al. 2010).

This section precises the results obtained from the literature review from the year 2007 to 2015 on the context of organizational factors that influences successful implementation of Green IT. The organizational factors were extracted from literature and include:

i.     firm or organizational size;
ii.    organizational climate and culture;
iii.   commitment, attitude and belief;
iv.    IT diffusion;
v.     competitive strategy and advantage;
vi.    green governance/leadership;
vii.   performance strategy; and
viii.  environmental impact of industry.

## 5.1 Organizational factors for Green IT implementation and Innovation Process

Business organizations play a critical role in reducing climate change and upholding environmental sustainability by using their dominance in the global economy (Melville 2010). As Melville explained "organizations pursue environmental sustainability by informing stakeholders of the need to make changes to business as usual, by motivating them to take actions to achieve environmental objectives, and by assessing the impact of such actions on economic and environmental performance" (Melville 2010, p. 14).

Environmental sustainability practices can lead to changes in organizational processes (Loeser et al. 2011). Due to the rapid growth in natural resource consumption coupled with the increase of greenhouse gas emissions; sustainability has become a crucial concept for corporate management (Lubin and Esty 2010). The top management team within organizations are looking for opportunities which are more innovative (Loeser et al. 2011).

The successful implementation of Green IT initiatives depends on several factors (Bose and Luo 2011). The head of management and IT managers should be aware of the factors influencing Green IT for successful implementation (Schmidt et al. 2010). By reviewing several literatures on organizational innovation and Green IT we intend to propose the following organization factors shown in the table below for successful implementation of Green IT as an administrative innovation(Damanpour 1996; Klein and Sorra 1996; Kuo and Dick 2010b; Opitz et al. 2014b) . We will also identify the relevance of these factors within the innovation processes such as idea generation, problem solving, implementation and diffusion (Utterback 1971).

Further in the following section we will discuss the importance of each factor on the context of the existing literature of Green IT. The strategy, design and planning of Green IT within an organization has recently become a pertinent topic in the IT literature (Bose and Luo 2011; Watson et al. 2010). By describing and reviewing the existing literatures we will attempt to provide pathway and assistance to the managers who are currently thinking of adopting Green IT within their organization and give them an idea about current observation on Green IT practices.

| References | Innovation Phase | Implementation Factors for Green IT |
|---|---|---|
| (Bose and Luo 2011); (Schmidt and Kolbe 2011) | Idea generation, Diffusion | Firms or Organizational size |
| (Opitz et al. 2014a); (Cooper and Molla 2012); (Molla 2008); (Molla et al. 2009a); (Vazquez et al. 2011); (Deng and Ji 2015) | Implementation, and Diffusion | Organizational Climate and culture |
| (Dalvi Esfahani, Abdul Rahman, & Zakaria, 2015);(Lei and Ngai 2014);(Opitz et al. 2014a);(Molla et al. 2014); (Molla et al. 2011); (Gholami et al. 2013) (Mithas et al. 2010) | Idea generation, Implementation, problem solving, and Diffusion | Commitment, Attitude and Belief |
| (Bose and Luo 2011); (Molla 2009a); (Molla et al. 2009b); (Deng and Ji 2015); (Chen et al. 2009) | Diffusion | IT diffusion |
| (Lei and Ngai 2014); (Erek et al. 2011); (Bose and Luo 2011); (Mann et al. 2009); (Vazquez et al. 2011); (Pollard 2015) | Idea generation | Competitive strategy and Advantage |
| (Lei and Ngai 2014); (Opitz et al. 2014a); (Cooper and Molla 2012); (Molla et al. 2014); (Bose and Luo 2011); (Loeser 2013); (Kuo and Dick 2010b); (Gholami et al. 2013); (Vykoukal 2010); (Molla 2009b); (Mann et al. 2009); (Sayeed and Gill 2009); (Butler and Daly 2009); Sarkar and Young (2009; (Mithas et al. 2010); (Molla et al. 2009b); (Tan et al. 2011) | Idea generation, Implementation, problem solving and Diffusion | Green Governance/ Leadership |
| (Opitz et al. 2014a); (Angeles and Fredericton 2015); (Schödwell et al. 2013); (Loeser 2013) | Implementation, problem solving and Diffusion | Performance Strategy |
| (Watson et al. 2010); (Elliot 2011); (Malhotra et al. 2013); (Molla et al. 2009a) | Implementation | Environmental Impact of Industry |

*Table 2: Organisational factors for Green IT implementation and innovation process*

## 5.2 Organizational size

Organizational size is an important factor in innovation implementation. According to Damanpour *"Organizational size would be more positively related to innovation in organizations operating under high than those operating under low environmental uncertainty"* (Damanpour 1996p. 696).

Larger firms possess greater potential for Green IT initiatives because the greater the number of employees the higher the awareness of Green IT (Schmidt and Kolbe 2011). Furthermore the more resources committed the greater the success rate of Green IT. Additionally the size of the organization is an important factor not only for Green IT implementation, it is also important for the innovation diffusion as well as innovation initiation (Bose and Luo 2011). According to Zhu large size organizations have more sufficient technical, financial, managerial resources to initiate Green IT (Zhu et al. 2006).

## 5.3 Organizational Climate and Culture

According to (Klein and Sorra 1996p. 1060) *"The more comprehensively and consistently implementation policies and practices are perceived by targeted employees to encourage, cultivate, and reward their use of a given innovation, the stronger the climate for implementation of that innovation"*. The organizational culture and climate has an important role in studying the relationship between IT adoption and diffusion.

The impact of organizational culture has a major impact on Green IT adoption (Deng and Ji 2015). Though IT plays a significant role in the protection of the environment, organizational employees' responsibilities have an important impact on Green IT adoption (Huang 2008). Organizations should be prepared to adopt new processes such as Green IT by developing new policy and procedures, business goals, systems and infrastructure and training (Vazquez et al. 2011), which is referred to as organizational readiness (Molla et al. 2009a). Organizational readiness also includes the awareness, and commitment of the employees as well as the readiness of suppliers, investors, partners and customers for Green IT (Molla et al. 2009a).

## 5.4 Commitment, Attitude and Belief

Employees' and senior managements' commitment towards Green IT is an important organizational factor. Earlier research proposes that successful IT implementation is possible when senior management is committed and includes the implementation process in wider approaches (Mithas et al. 2010).

Attitude towards Green IT is the reflection of the energy efficiency concerns in managing IT (Molla et al. 2011). Green IT attitude includes managerial attitude as well as employee attitude towards IT energy utilization (Molla et al. 2011). Individual commitment and acceptance of Green IT initiatives is important because the benefits will include reducing power consumption, lowering costs and sustaining a healthy life style (Dalvi Esfahani et al. 2015). According to Dalvi et al. the Green IT adoption studies focuses on recognizing the issues that influence individual's intention. In the IT adoption literature the effect of IT commonly represents attitude (Dalvi Esfahani et al. 2015).

According to Molla et al. (2014, attitudes towards Green IT refers to sentiments, values and norms of climate change, eco-sustainability and IT's role. Molla et al. (2014) also mentioned that the employees with positive attitudes towards Green IT are more likely to practice IT effectively. Furthermore according to Dalvi Esfahani et al. (2015) Green IT attitude positively impact their intension to practice IT pro-environmentally. Environmental belief is another key factor that influences environmental practice. According to Dalvi Esfahani et al. (2015) user's Green IT beliefs positively influence their attitude towards Green IT.

The green IT motivation and eco-efficiency motivation plays an important role for successful adoption of Green IT within a company (Molla and Abareshi 2011). Attitude towards Green IT is an affective characteristics of the senior managers which is imperative to Green IT success (Gholami et al. 2013).

## 5.5 IT Diffusion

The success of a Green IT innovation largely depends on how serious the organisation is at spreading or diffusing the innovation.

Having innovative ideas and diffusing them to related stakeholders are very significant for any organisation's success as they attain additional competitive advantage and market share according to the level of importance they give to innovations (Gunday et al. 2011). There are five attributes of innovation i) relative advantage; ii) compatibility; iii) complexity; iv) trialability and v) observability (Rogers Everett 1995). These attributes plays a crucial role in any organizational innovations that is Green IT (Van de Ven 1993).

Understanding the adoption and diffusion of green IS & IT across organizations informs the design of technological applications and institutional interventions to support ecological sustainability (Chen et al. 2009). There are some essential factors that influences the diffusion of Green IT (Molla et al. 2009b) such as: the maturity of green IT policies, practices, and technologies; Green IT governance, and Organizational motivations.

## 5.6 Competitive Strategy and Advantage

According to Bose and Luo (2011 "*organizations facing higher competition intensity are more likely to initiate Green IT*" (Bose and Luo 2011p. 49). As Lei and Ngai (2014 mentioned that higher level of apparent competitive advantage of Green IT will lead to higher level of intention to Green IT adoption.

Competitive advantage can result in reduced costs, reduced energy consumption as well as reduced waste discharge. All organizations seek profit, therefore the higher the environmental and economic benefit of Green IT the more likely top decision makers will decide to adopt Green IT (Lei and Ngai 2014). By implementing Green IT initiatives competitive advantage can be created (Erek et al. 2011). Pollard (2015) also described that Green IT has the possibility to construct fresh competitive opportunities. In addition Vazquez et al. (2011) described competitiveness as an element that is positively influenced by Green IT.

## 5.7 Green Governance/Leadership

The role of individual decision makers should not be ignored. Green IT is defined as a "*systematic application of ecological-sustainability criteria, such as pollution prevention, product stewardship, and use of clean technologies, for the creation, sourcing, use, and disposal of IT technical infrastructure, as well as within the human and managerial components of the IT infrastructure*" (Molla 2009bp. 757).

The top management of an organization can influence organizational policies and practices. They also play a major role on influencing employee behaviour (Mithas et al. 2010). Exploration of the role of IT manager is essential for Green IT adoption (Sarkar and Young 2009) which is Green governance. According to Molla et al. (2009b) *"Governance refers to the management infrastructure to implement Green IT. It is the operating model that defines the administration of Green IT initiatives to understand impacts, prioritise actions and manage an enterprise's responses"*.

According to Lei and Ngai (2014) there are many reasons driving organizations to accept Green IT, the ultimate choice on Green IT implementation is still made by organizational decision makers/top management (Lei and Ngai 2014).

The top manager's managerial understanding on environmental preservation is taken as a hazard or an prospect (Jackson and Dutton 1988). There are two aspects of leadership:

- First an employee, who is enforced by a passionate and dedicated leadership, has a better chance of surviving in the organization.
- Second the clear, positive and encouraging attitude of CIO's are crucial after the project is undertaken (Mann et al. 2009).

Sayeed and Gill (2009) suggested that Managerial process consists of management, incorporation, knowledge and reconfiguration. Furthermore, Lei and Ngai (2014said "*when environmental preservation is interpreted as an opportunity, higher level of personal norm on environmental preservation will lead to higher level of intention to Green IT adoption; when environmental preservation is interpreted as a threat, the relationship between personal norm on environmental preservation and intention to Green IT adoption will not be significant*". An Organization's leaders nurture the organizational shape as well as define the key issues to support Green IT initiatives (Kuo and Dick 2010a).

## 5.8 Performance Strategy

According to Huang (2008) only 35% of IT executives consider environmental issues, with the main reason for this negligence attributed to the inefficiency of organizations. Thus, for achieving Green IT implementation goals, organizations must have their own performance strategy. The top management must include an instrument on their policies to measure Green IT performance (Cooper and Molla 2012). The following technologies are used by some organizations for measuring their green IT performance -

- LCA – Life cycle assessment is a method which is used to assess the environmental influences of hardware design and manufacturing process (Angeles and Fredericton 2015).
- Web-based HP Carbon Footprint Calculator- This calculator help the employees of HP to assess their energy use and carbon emission as well as paper usage (Angeles and Fredericton 2015).
- ERT- Environmental reporting tool to generate notice to suppliers of material disclosure or reporting requirement.

## 5.9 Environmental impact of industry

The success of the Green IT initiative depends on the existing environmental laws. As global warming is a worldwide issue now governments are rewarding firms who are adopting green IT (Nishant et al. 2011). The challenge of climate change poses huge and widespread risk to people, societies, and the natural environment. The threat is real and immense and is ever increasing. This is the challenge of great urgency (Malhotra et al. 2013). According to United Nations survey environmental sustainability is the leading issue which will dominate the future.

# 6 CONCLUSION

We have analysed the existing literature of Green IT and proposed the phenomenon of Green IT as an administrative innovation and revealed the organizational factors of successful implementation. During our research we have noticed that there is no particular theory for Green IT (Bose and Luo 2011). Only very few of the studies explore the Green IT related issues after implementation (Tushi et al. 2014). The issues of Green IT after implementation and also before can be resolved by developing a theory for Green IT. As the environmental issue becomes more important every day, it is time for the IT community to be more aware of the problem and place emphasis on the promotion and adoption of sustainable IT usage and development approaches (Huang 2008).

For effective implementation of Green IT and to achieve the goal, learning about organizations information processes is required. Many papers have been published on Green IT (Loeser 2013) but limited work has pertained to the organizational factors for Green IT implementation. For the effectiveness of Green IT initiatives, organizational learning is necessary to improve organizational structure (Roome and Wijen 2006). The organizational factors included in this paper need to be considered for successful implementation of Green IT.

## 7  FUTURE RESEARCH

This research in progress paper intended to draw attention to the organizational factors for successful implementation of Green IT as an administrative innovation.  This research-in-progress paper has identified the existing gaps in the field of Green IT. After analysing 137 literatures for Green IT we can say that researchers are becoming more interested in this topic. This report tries to establish the phenomenon that if organizations implement Green IT as administrative innovation considering the organizational factors which includes organizational size, organizational climate/culture, IT diffusion, Commitment, Attitude and Belief, Competitive Strategy and Advantage, Green Governance/Leadership, Performance Strategy, Environmental impact of industry then Green IT implementation will be successful.

This paper did not include any case study to demonstrate Green IT as administrative innovation. After reviewing the existing publication we become aware that most of researchers used case study method. For future research we will conduct case studies on one successful organization and one unsuccessful organization that have implemented Green IT. Based on the case studies we endeavour to develop the followings:

- theory driven identification;
- process driven identification to see where the organization failed;
- Inter relationship between the organizational factors.

## Acknowledgements

We are grateful and thank the editors and the three reviewers for recommending our paper titled "Green Information Technology as Administrative innovation - Organizational factors for successful implementation: Literature Review" (paper 193) for publication and presentation at the 26th Australasian Conference on Information Systems (ACIS 2015).